\documentclass[journal=jctc,manuscript=article]{achemso}

\usepackage{graphicx}
\usepackage{dcolumn}
\usepackage{bm}

\usepackage[utf8]{inputenc}
\usepackage[T1]{fontenc}
\usepackage{mathptmx}
\usepackage{etoolbox}
\usepackage[normalem]{ulem}
\usepackage{mhchem}

\usepackage{color}

\usepackage{multirow}
\usepackage{booktabs}
\usepackage{pgfplotstable}
\usepackage{siunitx}

\usepackage{chemformula} 
\usepackage[T1]{fontenc} 
\usepackage{braket}
\usepackage{pdfpages}

\usetikzlibrary{arrows.meta,positioning,shapes.geometric,calc}

\author{Reza G. Shirazi}
\author{Alexander Zech}
\author{Peter Pinski}
\author{Vladimir V. Rybkin}
	\affiliation{HQS Quantum Simulations GmbH, Rintheimer Str. 23, D-76131 Karlsruhe, Germany}

\email{vladimir.rybkin@quantumsimulations.de}

\title[ASF benchmark]
{Performance of Automatic Active Space Selection for Electronic Excitation Energies}

\abbreviations{ASF, CASSCF, NEVPT2, MP2, DMRG}

\begin{document}
	
	
	\begin{abstract}
	Computation of electronic spectra is one of the most important applications of methods capturing static electron correlation, including complete-active-space self-consistent field (CASSCF) and post-CASSCF theories. Performance of these techniques critically depends on the active space construction, both in terms of accuracy and computational effort. In this work we benchmark the performance of automatic active space construction, as implemented in the Active Space Finder software, for the computation of electronic excitation energies. The multi-step procedure constructs meaningful molecular orbitals and selects the most suitable active space based on information from more approximate correlated calculations. It aims to tackle a key difficulty in computing excitation energies with CASSCF: choosing active spaces that are balanced for several electronic states. The Active Space Finder is tested with several established data sets of small and medium-sized molecules and shows encouraging results. We evaluate multiple setting configurations and provide practical recommendations.
	\end{abstract}
	
	\maketitle
	
\section{Introduction}
The complete active-space self-consistent-field (CASSCF) method and multireference approaches based on it are standard tools to capture static (strong) electron correlation \cite{roos1987complete}. Not always pronounced in the electronic ground states, static correlation is significantly more common in excited states, making multireference approaches an indispensable tool for electronic spectroscopy and photochemistry \cite{Lischka2018}. A well-known disadvantage of these techniques is the necessity to choose suitable orbitals for the active space. Traditionally, this is done manually, requiring insight into both the CASSCF method and the problem at hand, and introducing an element of subjectivity into such calculations.\cite{veryazov2011select} The active space size affects both accuracy and time-to-solution: in a straightforward implementation, CASSCF complexity scales exponentially with active space size. This puts a stiff limitation on the number of orbitals that can be included in the active space: increasing their number dramatically increases computational cost, not necessarily improving the accuracy of computed results. Thus, a "good" active space should be suitable to treat the problem at hand, but also sufficiently compact to maintain computational feasibility.

Already a non-trivial task for the ground state, active space construction becomes even more challenging when several states (ground and excited) are considered together, which is needed to compute electronic excitations either by state-averaged or state-specific CASSCF. Thus, development of reliable active space selection techniques can be critical to advance the broader application of multi-reference methods.

Up to now, several strategies for automatic active space selection have been developed. Early approaches were based on natural orbital occupation numbers \cite{pulay1988uhf, jensen1988second}. Recent developments along these lines have lead to adaptive basis sets for correlated wave functions (ABC family) \cite{Bao2018, Bao2019} and the active space selection based on 1st order perturbation theory (ASS1ST) \cite{Khedkar2019, Khedkar2020} scheme. Another philosophy makes use of quantum information measures: examples are autoCAS \cite{Stein2016, stein2019autocas}, which employs entropies to analyze orbital entanglement, and the quantum-information-assisted CAS optimization (QICAS)\cite{Ding2023} method. Alternative approaches to active space selection exploit projector/fragment-based techniques. Examples are atomic valence active spaces (AVAS) \cite{Sayfutyarova2017,   Claudino2019, Kolodzeiski2023, Lei2021}, automatic partition of orbital spaces based on singular-value decomposition (SPADE) \cite{Claudino2019, Kolodzeiski2023}, and imposed automatic selection and localization of CAS \cite{Lei2021}. Yet another family of methods utilizes ranking/scoring of orbitals particularly appealing for high-throughput calculations \cite{King2021, King2022}. Finally, machine learning techniques and data-driven approaches have found their way into active-space finding \cite{Jeong2020, Golub2021}. Of particular interest for the scope of this work are the methods extended to treat excited states: autoCAS, ABC2 and the dipole moment-based scheme \cite{Kaufold2023}.

Due to the importance of CASSCF and multireference methods for electronic spectroscopy \cite{Helmich_Paris2019}, their performance has been extensively studied; in some cases combined with automatic active space selection methods \cite{Hoyer2016, Kaufold2023, King2021, Bao2018}. This has resulted in readily available databases and datasets, including Thiel's set\cite{schreiber2008benchmarks} and the more recent QUESTDB database\cite{QUESTDB, QUESTDB2025}. The former encompasses both theoretical vertical and experimental values for excitation energies into several electronically excited states of 28 molecular systems. The latter is more extensive, contains several subsets \cite{QUESTDB1, QUESTDB2, QUESTDB3, QUESTDB4, QUESTDB6, QUESTDB7} with hundreds of compounds, and is subject to ongoing updates. Importantly, QUESTDB does not depend on experimental data by design, using high-level theory results as reference values.

Benchmarking multireference methods for electronic excitation energies has multiple facets: choosing the most suitable post-CASSCF electronic theory approach to calculate the dynamic correlation energy; benchmarking basis sets or method/basis combinations; and selecting the most suitable formalism: state-averaged \cite{Werner1981} or state-specific. Additional challenges arise if one is interested not only in vertical excitations, but in adiabatic ones and in vibronic structure. In this work, we restrict ourselves to benchmarking the automatic active space finding procedure within the state-averaged formalism for vertical electronic excitations. As a post-CASSCF method to calculate the dynamic part of the correlation energy we chose second-order n-electron valence state perturbation theory, NEVPT2 \cite{angeli2001introduction}. We utilize the strongly-contracted scheme for NEVPT2 (SC-NEVPT2) \cite{ANGELI2001297}, which has been shown to systematically deliver fairly reliable vertical transition energies \cite{Sarkar2022}, only marginally inferior to the partially contracted scheme \cite{Angeli2002}. While not exhaustive in scope, this approach provides a systematic and unambiguous test for active space construction for several electronic states.

%
%
\section{Methods}

\subsection{Requirements for automatic active space determination}
We believe it is desirable that a generally applicable, user-friendly and computationally
affordable technique to find active spaces satisfies the following criteria:
\begin{enumerate}
	\item it generates orbitals that serve as a good guess for CASSCF, leading to convergence within a reasonable number of iterations;
	\item the active space is constructed through an automatic procedure that minimizes the need for manual user intervention and maximizes reproducibility;
	\item autonomy: in line with the \textit{ab initio} character of multi-reference methods,
	the choice is made independently of problem-specific reference data (even though the active space selection procedure may involve parameters of a generic nature);
	\item \textit{a priori} character: the choice is made prior to the CASSCF calculation.
\end{enumerate} 
At first sight, the final point might appear as a tautology. However, active spaces can be refined
by repeating CASSCF calculations with different numbers of active orbitals, especially, if the
molecular orbitals from one calculation are taken as a guess for the next one. This approach is not
only employed in manual active space selection, but is also used by some automatic schemes.\cite{Khedkar2019,Kaufold2023}
The philosophy underlying this work is that the active space should be determined prior to any CASSCF
or post-CASSCF calculations, even if they are inexpensive as in ref.~\cite{Shirazi2024}. This permits
the approach to be applied in a more general context to large active spaces, for which calculations
may be expensive. It also makes the scheme suitable for methods that do not optimize orbitals, such as complete active space
configuration interaction (CASCI).
\subsection{Algorithm}

This work employs the Active Space Finder (ASF) package,\cite{ASF_github} which is published
together with its documentation under an open source license.
Modifications of the software were made for an improved treatment of excited
states as part of this work. Those changes will be released in the publically available software
repository in due course. A detailed description of the principles underlying the ASF software,
with a focus on ground states, is in preparation for a forthcoming manuscript. Therefore, the
algorithm is explained below only in general terms, and a description of the modifications made for
excited states is provided.

The central component of our active space finding procedure is a density matrix renormalization
group (DMRG)\cite{chan2011density} calculation performed with low-accuracy settings.
This idea is related to the autoCAS method by Reiher and co-workers.\cite{Stein2016} However, the
analysis of the DMRG results to determine an active space employs profoundly different principles.
Prior to performing the DMRG calculation, an initial active space needs to be selected.
It has to be sufficiently large, such that it contains all orbitals that will be part of the final
active space. At the same time, it must be sufficiently small to keep the effort for the DMRG
calculation feasible. The individual steps are described in more detail below.

\begin{description}

	\item[Self-consistent field calculation] as the first step.
	The fully automatic mode of the ASF	always employs the spin-unrestricted Hartree-Fock (UHF)
	method. A stability analysis is performed with the converged orbitals, followed by a restart of
	the calculation if there is an internal instability.
	Pulay and co-workers demonstrated the utility of unrestricted natural orbitals
	(UNOs) for the selection of minimal active spaces.\cite{pulay1988uhf,pulay2015unos}
	We do not calculate UNOs, but we employ symmetry breaking in a similar spirit. 
	Therefore, UHF calculations are initiated even if the system has singlet multiplicity.
	However, it is also possible to carry out a restricted Hartree-Fock (RHF) calculation manually
	and to use it to proceed further with the ASF software.

	\item[Selection of an initial space.] Even relatively small molecules may possess hundreds of
	occupied and virtual molecular orbitals with a basis set suited for correlated calculations. An
	initial active space, which may be substantially larger than the final space, needs to be
	determined in order to keep the computational effort manageable for the subsequent steps. All
	procedures used in the present work employ natural orbitals of an orbital-unrelaxed
	second-order M\o{}ller-Plesset perturbation theory (MP2) density matrix for the ground state.
	With the help of an occupation number threshold, an initial set of orbitals is selected, and
	further orbitals are discarded if their number exceeds an upper limit. It is important to omit
	orbital relaxation effects, as these lead to artifacts with unphysical eigenvalues of the
	density matrix below zero or above two. A density-fitting MP2 implementation is used for the
	sake of efficiency. It is particularly beneficial to employ the UHF solution from the first
	step as a reference wave function for the MP2 calculation. We tested two different ways to
	represent the initial active space:

	\begin{enumerate}
		\item In the first variant, the MP2 natural orbitals are propagated to the next step.
		Therefore, the finally chosen active space will be a subset of the MP2 natural orbitals
		selected for the initial active space.
		\item Ground state MP2 natural orbitals are not necessarily the most appropriate orbital
		set for excited electronic states. Therefore, we implemented a re-canonicalization of the
		initial space in a procedure that is analogous to quasi-restricted orbitals
		(QROs).\cite{Neese2006} In contrast to the QRO procedure, the respective Fock matrix
		sub-blocks are projected to the initial active space constructed from MP2 natural orbitals
		prior to diagonalization. This ensures that the initial space is preserved without mixing
		in other orbitals, even though it is represented by orbitals resembling a restricted
		open-shell Hartree-Fock (ROHF) solution. In contrast to using ROHF orbitals or QROs
		directly, our procedure provides a clear-cut selection of a sufficiently, but manageably
		large initial active space through the occupation number threshold for MP2 natural
		orbitals.
	\end{enumerate}

	\item[Approximate CASCI calculation] to determine correlation information for the final active
	space selection. This step is carried out as a DMRG-CASCI calculation with a low
	matrix product state bond dimension.
	If the size of the initial space is sufficiently small (typically up to 14 orbitals), a
	regular CASCI calculation can be performed instead. It is important to point out that the
	nature of the initial space selection determines the type of molecular orbitals used in the
	(DMRG-)CASCI calculation and for the final active space selection. Thus the molecular orbitals
	used throughout the calculation will be MP2 natural orbitals, QROs, or another basis, depending
	on the initial choice.
	
	\item[Analysis of the correlation information] from the (DMRG-)CASCI calculation. An important
	design goal of the ASF was to identify correlation partner orbitals. This is accomplished
	through an automatic analysis of the two-electron cumulant. It has previously been discussed in
	the literature as a measure for electron correlation,\cite{Kong2011,Hanauer2012}
	but the ASF is, to the best of our knowledge, the first method employing it for active space
	selection. In contrast to mutual information calculated from two-orbital
	entropies,\cite{Stein2016} which include contributions of reduced density matrix elements of up
	to fourth order,\cite{Boguslawski2015} only the two-electron density matrix is required to
	compute the two-electron cumulant.

	Analysis of the cumulant leads to multiple suggestions for sensible active spaces of different
	sizes. In order to select one of these active spaces, the one-orbital entropy is used as an
	auxiliary criterion.\cite{Boguslawski2012,Boguslawski2015}
	An entropy threshold is defined as a target value. For all active space suggestions identified
	through analysis of the cumulant, the lowest entropy of all orbitals in the active space is
	compared to the target value. The active space with its lowest entropy closest to the target
	value is selected as the final choice.

\end{description}

Extensions of the aforementioned algorithm were made in order to construct active spaces suitable
for the treatment of excited states with state-averaged CASSCF. We implemented two strategies:

\begin{description}

    \item[Union of individual active spaces.] The (DMRG-)CASCI density matrices are calculated for
    all relevant electronic states. Subsequently, the correlation information for each electronic
	state is analyzed to determine one space per electronic state independently.
	Finally, a combined active space is created as the union: every orbital that is member of an
	active space of at least one electronic state becomes a member of the combined space.

	\item[Averaging of electronic states.] Instead of forming the union of multiple active spaces
	determined for each state individually, active spaces are determined for multiple states at
	once. For this purpose, averages of the cumulant of the one-orbital density are formed.
	We define the two-electron cumulant ${}^2_n\lambda^{pq}_{rs}$ for a single electronic state $n$
	in terms of the respective reduced density matrices:
	\begin{multline}
	  {}^2_n\lambda^{pq}_{rs}
	    = \\\braket{\Psi_n | \hat{a}^{\dagger}_p \hat{a}^{\dagger}_q \hat{a}_s \hat{a}_r | \Psi_n}
		- \braket{\Psi_n | \hat{a}^{\dagger}_p \hat{a}_r | \Psi_n}
		\braket{\Psi_n | \hat{a}^{\dagger}_q \hat{a}_s | \Psi_n}
		+ \braket{\Psi_n | \hat{a}^{\dagger}_p \hat{a}_s | \Psi_n}
		\braket{\Psi_n | \hat{a}^{\dagger}_q \hat{a}_r | \Psi_n}.
	\end{multline}
	Here, $p,q,r,s$ refer to spin orbitals, while $\Psi_n$ represent the CASCI wave functions of
	the respective electronic state. An averaged cumulant is constructed for multiple electronic
	states $N_\text{states}$, assuming equal weighting:
	\begin{equation}
	    {}^2\overline{\lambda}^{pq}_{rs} = N^{-1}_\text{states} \sum_n {}^2_n\lambda^{pq}_{rs}
	\end{equation}
	Instead of analyzing the cumulant of each state individually, the algorithm is applied directly
	to ${}^2\overline{\lambda}^{pq}_{rs}$. Multiple suggestions for active spaces of different
	sizes are determined for state-averaged calculations analogously as for ground states.

	In order to select a single space, entropies are determined from an averaged one-orbital
	density:
	\begin{equation}
	    s_i = -\sum_\alpha^{-,\uparrow,\downarrow,\uparrow\downarrow}
		\overline{\omega}_{i,\alpha} \ln \overline{\omega}_{i,\alpha}
	\end{equation}
	The averaged one-orbital density $\overline{\omega}_{i,\alpha}$ for each spatial orbital $i$ is
	calculated using the respective one-orbital density $\omega^n_{i,\alpha}$ of each state $n$:
	\begin{equation}
	    \overline{\omega}_{i,\alpha} =
		N^{-1}_\text{states} \sum_n^{N_\text{states}} \omega_{i,\alpha}
	\end{equation}

\end{description}

The same set of orbitals must be used for each electronic state; this is  particularly important to
recognize if the active space is chosen for more than one spin state. Often, it will be
advantageous to perform the Hartree-Fock calculation, and the subsequent MP2 calculation (if
applicable), for the state with the highest spin multiplicity. That single set of orbitals can be
used for DMRG-CASCI calculations for different spin states to determine the final active space.

For technical reasons, active space selection via averaging of the cumulant and of the one-orbital
density have been implemented for electronic states of the same multiplicity only. When
determining an active space for state-averaged CASSCF calculations with more than one
spin state, it is still possible to employ the averaging procedure for each multiplicity
separately; however, the final active space is obtained as the union of the spaces determined with
averaging for each respective spin-state only. In principle, it would be conceivable to extend the
procedure such that it averages the relevant quantities over multiple spin states using appropriate
weighting factors.

\subsection{Details of the ASF Calculation}



Several calculation variants were tested by altering the following parameters:
\begin{itemize}

  \item  Entropy Threshold: the default value for the entropy threshold in the ASF is set to $0.1 *
  log(4) \approx 0.14$. We also tested the selection of larger active spaces obtained with a
  somewhat lower entropy threshold of $0.11$. Calculations utilizing the low entropy threshold are
  denoted as ``l-ASF'', whereas ``ASF'' stands for the default threshold.
  \item Spin in the HF (SCF) calculation: in some scenarios, it is beneficial to generate guess
  orbitals for excited states through an SCF calculation with triplet spin multiplicity. As a
  result, MP2 natural orbitals are also calculated for the lowest triplet state using the UHF
  solution as the reference. After truncation to an initial active space, (DMRG-)CASCI calculations
  and the subsequent active space selection steps are performed for the respective singlet states;
  only the molecular orbital coefficients originate from a triplet state calculation. Such
  calculations, employing triplet state guess orbitals, are denoted with ``ASF(T)''. Calculations
  performed with singlet spin multiplicity throughout are indicated by ``ASF(S)''.
  \item Rotation of triplet-state orbitals: UHF-MP2 natural orbitals are often an excellent guess
  for ground-state CASSCF calculations. However, for excited states it is often preferable to
  employ orbitals with shapes holding closer semblance to a canonical Hartree-Fock solution. For
  this reason, the ASF provides functionality to transform the initial active space to a QRO-like
  solution, without mixing between the initial active and inactive spaces. We performed
  calculations where the orbitals were obtained using UHF and MP2 calculations for the triplet
  state, as in ASF(T), but subsequently rotated with the QRO transformation. These calculations are
  referred with QRO in parantheses: ``ASF(QRO)''.
  \item Calculations reported in this work determined active spaces by averaging cumulants and
  one-orbital densities over the lowest two singlet states, and analyzing the averaged quantities.
  Reporting results with unions of active spaces determined individually for each state will be
  left to future work.
  \item In order to avoid accidental convergence to higher states, the (DMRG-)CASCI calculations
  for active space selection were performed for the lowest four electronic states. However, active
  spaces were determined only for the lowest two singlet states, with the cumulant and the
  one-orbital density averaged only over these two states.
    
 \end{itemize}

The various simulation options are summarized in Table \ref{tab:acronyms}. 

\begin{table}
	\begin{tabular}{|l|l|l|l|}
		\hline
		 Entropy threshold = 0.14 & ASF(S) & ASF(T) & ASF(QRO) \\
		 \hline
		 Entropy threshold = 0.11 & l-ASF(S) & l-ASF(T) & l-ASF(QRO) \\
		\hline\hline
		Spin in SCF & S = 1 & S = 3 & S = 3 \\
		\hline
		Initial orbitals & MP2(S = 1) & MP2(S = 3) & MP2(S = 3) \\
		\hline
		Rotation on initial active space & natural & natural & QRO \\
		\hline
		Spin in ASF & S=1 & S=1 & S=1 \\
		\hline
	\end{tabular}

	\caption{{\label{tab:acronyms}}Acronyms for active-space-finding schemes with basic properties.}
\end{table}

\subsection{Computational set-up}

This work employs the \textit{Active Space Finder (ASF)} package \cite{ASF_github}, which is developed in Python, and built on top of the \textit{PySCF} \cite{Sun2018pyscf} and \textit{Block2} \cite{Zhai2023} packages. The code changes for excited states, which are subject to testing in this work, were implemented in a development version of the ASF that will be made available publically in due course.
PySCF was used for all quantum chemistry calculations, including CASSCF \cite{Sun2017CASSCF} and SC-NEVPT2 \cite{Angeli2002}. Block2\cite{Zhai2021} was employed via its PySCF interface for DMRG calculations called by the ASF.

The triple-zeta quality basis set with diffuse functions def2-TZVPD \cite{Weigend2005, Rappoport2010} was used for all calculations. Resolution-of-identity (or density fitting) \cite{VAHTRAS1993} was applied to accelerate electron repulsion integral evaluation using the corresponding auxiliary basis sets\cite{Weigend2008}. Excitation energies were computed using state-averaged CASSCF with equal weights for the ground and the first excited state. Dynamic correlation energy contributions were calculated using the strongly contracted NEVPT2 formalism.

This work focuses on vertical excitation energies between the ground and the first excited state, which were computed using accurate ground state equilibrium geometries taken from literature.

\subsection{Testing data sets}
Testing was performed with 32 small and medium-sized molecules from Thiel's data set\cite{schreiber2008benchmarks} and from the paper of Hoyer \textit{et al} \cite{Hoyer2016}. This work employs molecular geometries and reference excitation energies as described in those articles. For molecules that are present in both papers, we chose the data from the more recent work in ref.~\citenum{Hoyer2016}. The complete list of molecules is available in the SI.



\section{Results and Discussion}

\begin{figure}
	\includegraphics[scale=0.85]{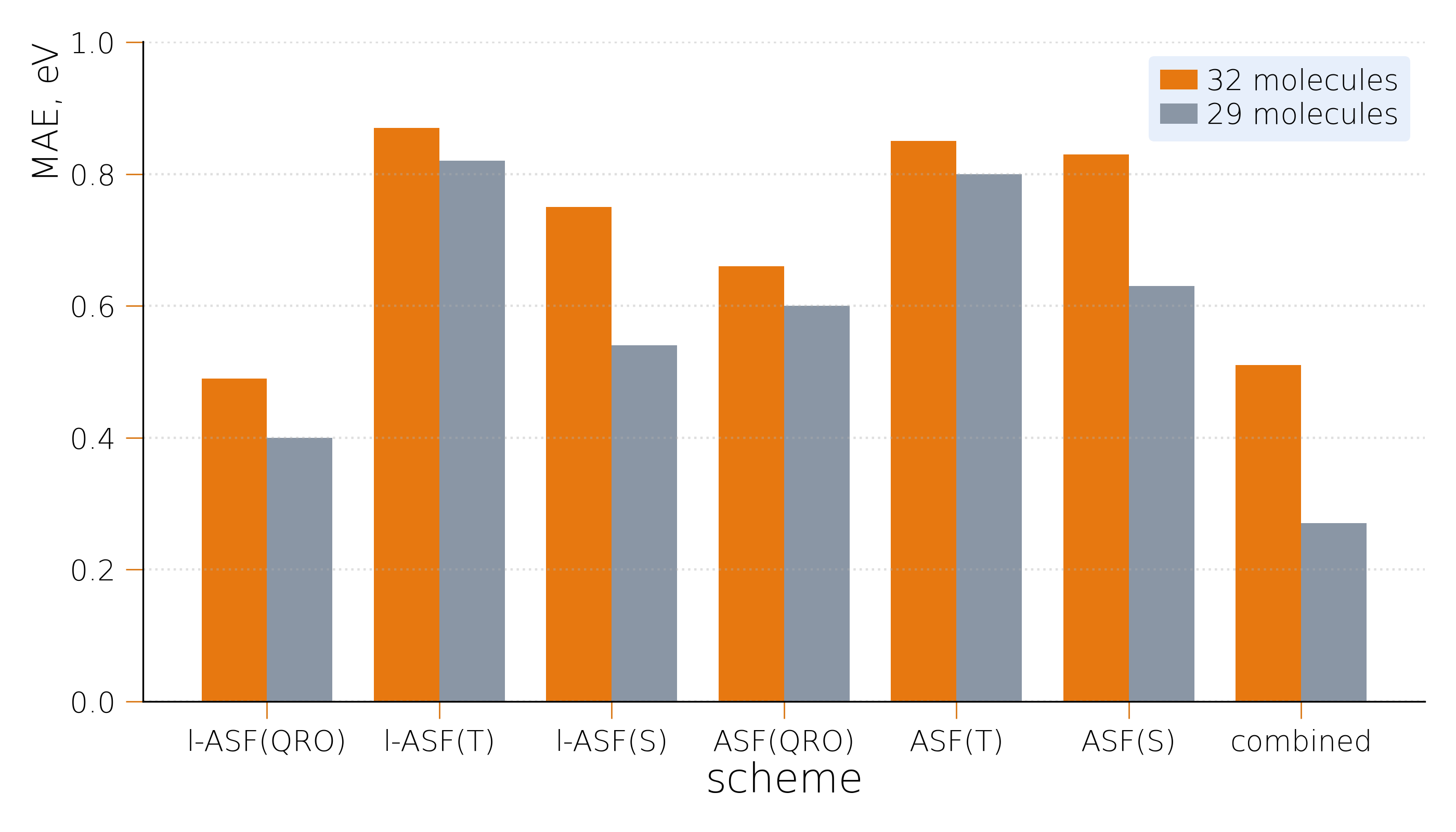}
	
	\caption{\label{fig:MAE} Mean absolute errors wrt. benchmark excitation energies values for the test data set. Orange - the entire data set of 32 molecules; grey - reduced data set (29 molecules).}
\end{figure}

The results of calculations using ASF are summarized in Table \ref{tab:results} and Figure \ref{fig:MAE}, whereas the full data are given in the SI.
Active space selection is not a problem with a uniquely defined solution: depending on the problem at hand and the electronic structure method employed, active spaces of different size may be suitable for the same molecule.
This work takes a pragmatic approach: it examines how well automatic active space selection and correlated calculations work together in reproducing reference excitation energies.

We assign calculation results into three categories: failure (CASSCF has not reached convergence), miss (mean absolute deviation, MAE, from the reference value is larger than 1 eV) and satisfactory performance.
If CASSCF fails to converge, which is typically a clear sign of improper active-space selection, the deviation is assigned the value of the reference excitation energy.

\begin{table}\label{main_stats}
	\begin{tabular}{|l|l|l|l|l|l|l|l|}
	\hline
	method& l-ASF(QRO)  & l-ASF(S)  & l-ASF(T) & ASF(QRO) & ASF(S) & ASF(T) & combined \\
	\hline
	fail & 0 & 3 & 2 & 1 & 2 & 2 & 1 \\
	miss	& 7 & 3 & 5 & 8& 8 & 8 & 3\\
	MAE	& 0.49 & 0.75 & 0.87 & 0.67 & 0.83 & 0.85 & 0.51 \\
	\hline 
	\end{tabular}
	\caption{{\label{tab:results}}Main benchmarking results: "fail" represents the number of systems with CASSCF convergence failure; "miss" denotes the number of systems with absolute deviation from the reference larger than 1 eV; MAE is the mean average error with respect to reference values for the entire data set of 32 molecules.}
\end{table}

 First, we note that for any of the tested schemes number of failures does not exceed 3 out of 32, whereas the maximum number of misses and failures is 10, just below one third. MAE for the entire dataset is below 1 eV for all schemes. For both values of entropy thresholds, rotating the initially selected MP2 natural orbitals with the QRO transformation leads to improved performance. The best results are obtained with the ``l-ASF(QRO)'' scheme, which combines triplet state guess orbitals with a QRO transformation and a lowered entropy threshold: with these active spaces, all CASSCF calculations converged giving only eight misses, whereas MAE is slightly above 0.5 eV. 

It is useful to understand the reasons behind failures and misses to be able to propose a diagnostic for the sanity of results without comparing to the reference. One cause of unsatisfactory performance is the presence of a near-degenerate excited state not considered by state averaging. This can lead to root-swapping in the CASSCF procedure. Additionally, some perturbation theories for dynamic correlation are prone to intruder states. Para-benzoqinone is a notorious example: its first and second singlet excited states are separated by ca. 0.2 eV according to the experiments (and less than 0.1 eV depending on the theory level). \cite{schreiber2008benchmarks} Therefore, our set-up considering  ground and first excited state only leads to either miss or failure in all applied schemes. One can remedy this problem by inspecting excited states at Step 3 of the algorithm: fast DMRG calculation can include more excited states than needed in the production run to estimate the possibility of near-degeneracy. Although DMABN molecule does not have such near-degenerate levels, averaging over more states is known to increase the accuracy of multireference calculations of its electronic spectra critically. \cite{DMABN} 

An additional reason for unsatisfactory performance is that larger active spaces are needed than
those determined by the ASF with default settings. For propanamide, all protocols result find an
(4,4) active space, result in failures and misses. Analysis of the active orbitals shows that the
calculations converge to an electronic state different from the true first excited $n \rightarrow
\pi^*$ one. This is a sign that more states must be considered for state averaging. In the case of
cyclopentadiene, the only protocol delivering accurate results (error of just 0.13 eV) is l-ASF(S),
which determines an (8,8) active space, whereas other schemes suggest (4,4) active spaces and
perform unsatisfactorily. For formamide, not only a (4,4) but also a (6,6) active space lead to
significant errors; only ASF(S) and l-ASF(S) lead to accurate values, as they determine larger
(8,9) active spaces.  Therefore, we recommend to avoid production calculations if a small active
space is generated by the selected scheme. Instead, active-space construction should be
reconsidered with either the orbital entropy threshold relaxed, or with a different choice of
initial orbitals.

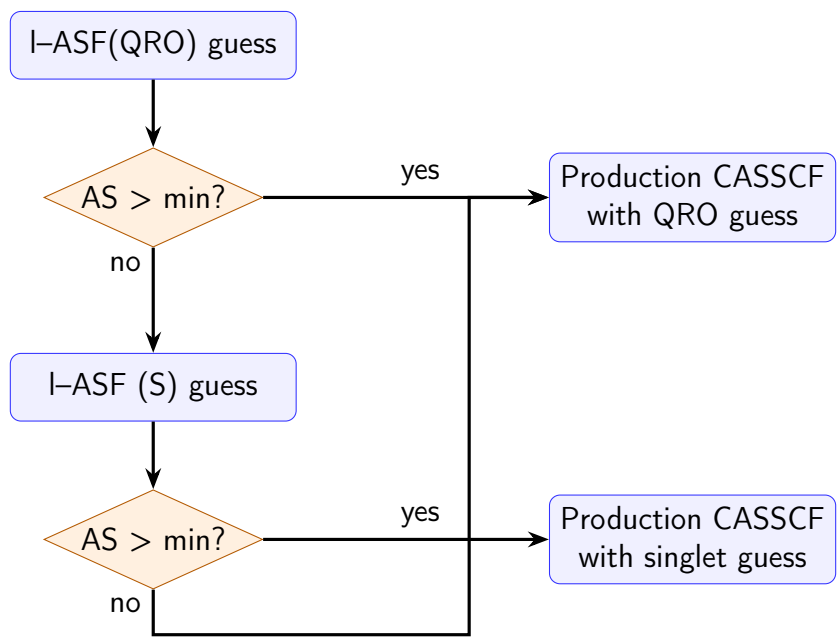
\begin{figure}[t]
	\centering
\begin{tikzpicture}[
	process/.style={draw=blue!80, fill=blue!6, rounded corners, minimum width=38mm, minimum height=9mm, align=center},
	decision/.style={diamond, draw=orange!70!black, fill=orange!12, aspect=2.2, align=center, inner sep=1.2pt},
	>={Stealth[length=2.8mm]},
	line/.style={->, line width=1.2pt},
	node distance=9mm and 16mm,
	font=\sffamily
	]
	
	\node[process]  (g1) {l–ASF(QRO) guess};
	\node[decision, below=of g1] (d1) {AS > min?};
	\node[process, right=38mm of d1] (p1) {Production CASSCF\\with QRO guess};
	
	\node[process, below=14mm of d1] (g2) {l–ASF (S) guess};
	\node[decision, below=of g2] (d2) {AS > min?};
	\node[process, right=38mm of d2] (p2) {Production CASSCF\\with singlet guess};
	
	
	\draw[line] (g1) -- (d1);
	\draw[line] (d1) -- node[above, pos=0.55] {yes} (p1.west);
	\draw[line] (d1.south) -- ++(0,-7mm) node[left, pos=0.3] {no} -- (g2.north);
	
	\draw[line] (g2) -- (d2);
	\draw[line] (d2) -- node[above, pos=0.55] {yes} (p2.west);
	
	\draw[line] (d2.south) -- ++(0,-6mm) node[left, pos=0.35] {no} -- ++(42mm,0) |- (p1.west);
	
\end{tikzpicture}	
	\caption{A combined ASF protocol: one loops over guess orbitals generation methods until the active space has more than three virtual orbitals, which is denoted as \textit{min} in the diagram. The order of the methods is: 1) QRO, 2) singlet (S) based on their performance. In rare case where both methods suggest a small active space, a QRO guess should be used.}
	\label{fig:workflow-combined}
\end{figure}

Inspired by these results, we have designed a combined protocol shown in Figure
\ref{fig:workflow-combined}, which addresses the issue of small active spaces. The first step is an
l-ASF(QRO) calculation. If the active space thereby generated contains fewer then three unoccupied
orbitals, l-ASF(S) is used, instead. If this scheme results in a small active space, too, one
should perform a production CASSCF calculation with an active space from l-ASF(QRO). The only and
notorious exception is the water molecule, for which a small (4,5) active space is balanced and
provided very accurate excitation energies, whereas a larger (6,6) active space from l-ASF(S) leads
to larger errors. Despite its importance, water is a small system and it is not representative. 

To get a clearer picture, we have considered a reduced data set by removing molecules requiring averaging over more than two states: propanamide, benzoquinone and DMABN. The performance of all protocols is considerably improved, especially for the combined scheme \ref{fig:MAE}. Independently of this, we have been able to achieve accurate results for p-benzoquinone and propanamide by averaging over four electronic states.  However, consideration of multiple electronic excitations goes beyond the scope of the current work.

\section{Conclusions}

We have designed and tested automatic active space selection protocols to compute excitation
energies with multi-reference methods. To this end, we worked with the publically available
``Active Space Finder'' (ASF)\cite{ASF_github} software. Employing multiple levels of correlated
calculations, it constructs suitable orbitals and selects active spaces with the help of the
two-electron cumulant and one-orbital entropies. As part of this work, the implementation was
adapted for an improved treatment of excited states targeting state-averaged CASSCF.

Testing its performance for electronic excitation energies of a set of small and medium-sized organic molecules reveals reasonable performance: with all schemes employed the percentage of unsatisfactory results varies from 25 to 30\% in fully automatic mode. The best-performing scheme utilizes a slightly lowered entropy threshold for active orbital selection and is based on quasi-restricted orbitals (QRO) transformed within an initial space constructed from MP2 natural orbitals. 

Error analysis has revealed two main sources of poor performance: the presence of quasi-degenerate excited states not considered in the calculation and small sizes of active spaces. Both can be remedied by averaging over more excited states, repeating the preliminary calculations with a different orbital entropy criterion and/or alternative choice of initial orbitals. Such modifications are easy to customize and can be readily streamlined for high-throughput calculations. We have proposed a simple combined workflow addressing the problem of small active sizes, which significantly reduces the errors.

\section{Acknowledgements}
This work has received funding from the German Federal Ministry of Research, Technology and Space (BMFTR) within the PhoQuant project (Grant No. 13N16103) and the PASQUOPS project (Grant No. 13N17251).
The Active Space Finder software was developed in collaboration with and with support by Covestro Deutschland AG.

	\bibliography{ASF_excited.bib}

	
\appendix
\section*{Supporting Information}
\addcontentsline{toc}{section}{Supplementary Information} 

\includepdf[pages=-]{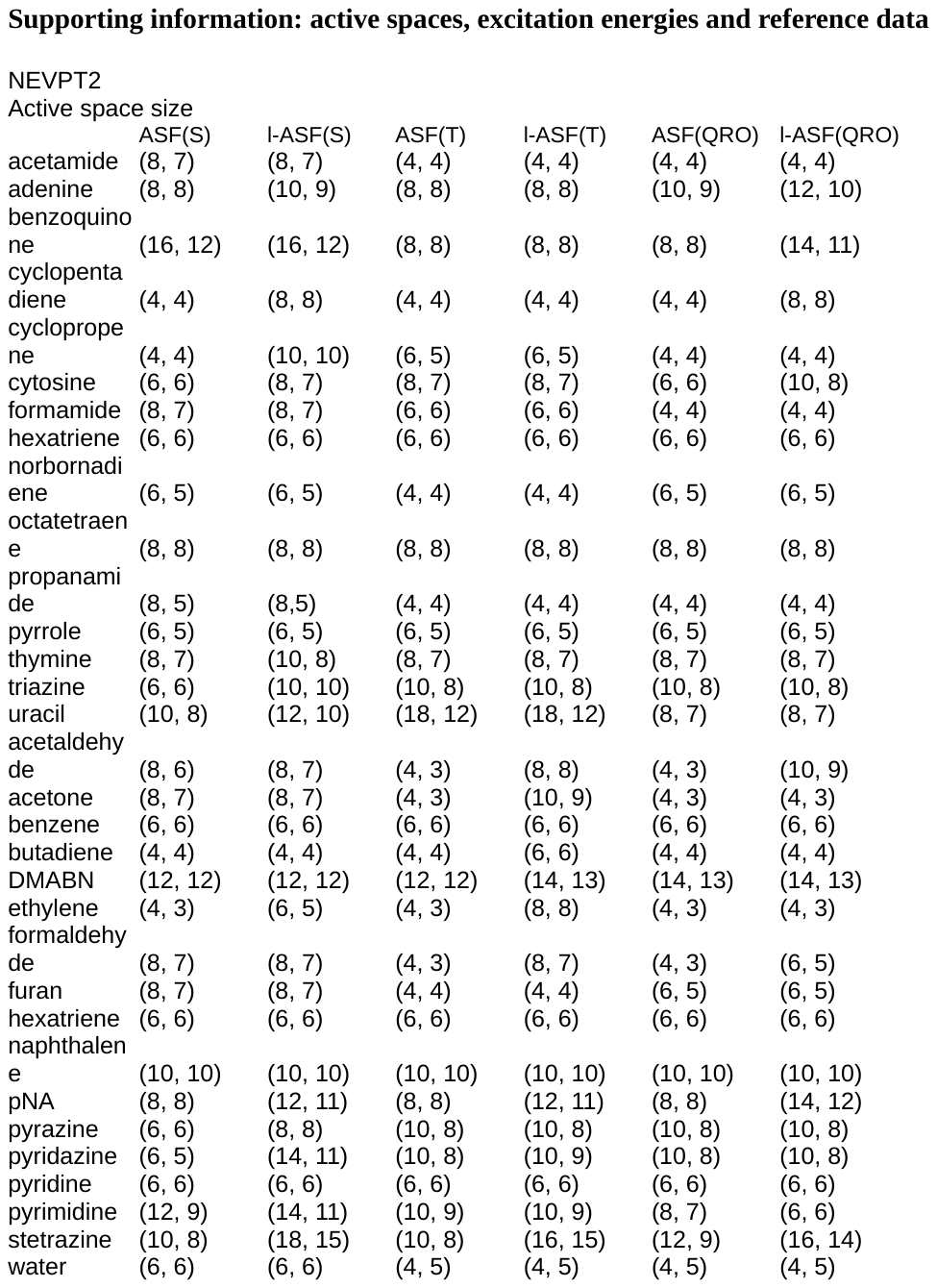} 

\end{document}